\documentclass[prb,twocolumn,a4paper,superscriptaddress,floatfix,showpacs]{revtex4}
\usepackage[english]{babel}
\usepackage{graphicx}
\usepackage{subfigure}
\usepackage{color}
 \expandafter\let\csname equation*\endcsname\relax
  \expandafter\let\csname endequation*\endcsname\relax
\usepackage{amsmath,amssymb,amsfonts}

\begin{document}

\title[Effect of Eu doping and partial oxygen isotope
substitution]{Effect of Eu doping and partial oxygen isotope
substitution on magnetic phase transitions in
(Pr$_{1-y}$Eu$_y$)$_{0.7}$Ca$_{0.3}$CoO$_3$ cobaltites}

\author{N.~A.~Babushkina} \email[e-mail:] {babushkina_NA@nrcki.ru}
\affiliation{National Research Center
``Kurchatov Institute'', Kurchatov Square 1, Moscow, 123182
Russia}

\author{A.~N.~Taldenkov} \affiliation{National Research Center
``Kurchatov Institute'', Kurchatov Square 1, Moscow, 123182
Russia}

\author{S.~V.~Streltsov} \affiliation{Institute of Metal Physics,
Ural Branch, Russian Academy of Sciences, S. Kovalevskaya Str. 18,
Ekaterinburg, 620990 Russia} \affiliation{Ural Federal University,
Mira Str.  19, Ekaterinburg, 620002 Russia}

\author{A.~V.~Kalinov} \affiliation{All-Russian Electrical
Engineering  Institute, Krasnokazarmennaya Str.\ 12, 111250
Moscow, Russia}

\author{T.~G.~Kuzmova} \affiliation{Department of Chemistry,
Moscow State University, 119991 Moscow, Russia}

\author{A.~A.~Kamenev} \affiliation{Department of Chemistry,
Moscow State University, 119991 Moscow, Russia}

\author{A.~R.~Kaul} \affiliation{Department of Chemistry, Moscow
State University, 119991 Moscow, Russia}

\author{D.~I.~Khomskii} \affiliation{$II.$ Physikalisches
Institut, Universit\"at zu K\"oln, Z\"ulpicher Str. 77, 50937
K\"oln, Germany}

\author{K.~I.~Kugel} \affiliation{Institute for Theoretical and
Applied Electrodynamics, Russian Academy of Sciences, Izhorskaya
Str. 13, Moscow, 125412 Russia}

\date{\today}

\begin{abstract} We study experimentally and theoretically the
effect of Eu doping and partial oxygen isotope substitution on the
transport and magnetic characteristics and spin-state transitions
in (Pr$_{1-y}$Eu$_y$)$_{0.7}$Ca$_{0.3}$CoO$_3$ cobaltites. The Eu
doping level $y$ is chosen in the range of the phase diagram near
the crossover between the ferromagnetic and  spin-state
transitions ($0.10 < y < 0.20$).  We prepared a series of samples
with different degrees of enrichment by the heavy oxygen
isotope $^{18}$O, namely, containing 90\%, 67\%, 43 \%, 17 \%, and
0 \% of $^{18}$O. Based on the measurements of ac magnetic
susceptibility $\chi(T)$ and electrical resistivity $\rho(T)$, we
analyze the evolution of the sample properties with the change of Eu
and $^{18}$O content. It is demonstrated that the effect of
increasing $^{18}$O content on the system is similar to that of
increasing the Eu content. The band structure calculations of the
energy gap between $t_{2g}$ and $e_g$ bands including the
renormalization of this gap due to the electron-phonon interaction
reveal the physical mechanisms underlying such similarity.
\end{abstract}

\pacs{
72.80.Ga,\,
75.30.Wx,\,
32.10.Bi,\,
71.27.+a,\,
71.15.Mb,\,
75.25.Dk 
}

\maketitle

\section{Introduction \label{intro}}

Most magnetic oxides are characterized by a strong interplay of
electron, lattice, and spin degrees of freedom giving rise to
multiple phase transitions and different types of ordering. The
phase transitions are often accompanied by the formation of
different inhomogeneous states. In such a situation, the oxygen
isotope substitution provides a unique tool for investigating
inhomogeneous states in magnetic oxides, which allows studying the
evolution of their properties in a wide range of the phase
diagram. Sometimes, particularly if a system is close to the
crossover between different states (usually leading to a phase
separation), the isotope substitution can lead to significant
changes in the ground state of the system~\cite{Babushkina1998}.

A good example of  such phenomena is provided by cobaltites. These
perovskite cobalt oxides  have attracted a special interest owing
to the possibility of the spin-state transitions (SST) for the Co
ions induced by temperature or
doping~\cite{goodenough.jap65,asai.jpsj98,saitoh.prb97,
tokura.prb98,korotin.prb96,berggold.prb08,ovchinnikov.UFN10} and
the related phase separation
phenomena~\cite{maignan.jpcm02,leighton.prb03,
louca.prl06,podlesnyak.prl08,sboychakov.prb09,
louca.prb09,leighton.prb10,podlesnyak.prb11}. The effect of
$^{16}$O $\rightarrow ^{18}$O isotope substitution on the
properties of (Pr$_{1-y}$Eu$_y$)$_{0.7}$Ca$_{0.3}$CoO$_3$
cobaltites ($0.12 <y < 0.26$) was studied earlier in our paper
\cite{Kalinov2010}. It was found that with increasing Eu content,
the ground state of the compound changes from a ``nearly metallic"
ferromagnet (ferromagnetic metallic clusters embedded into an
insulating host) to a ``weakly magnetic insulator" at $y < y_{cr}
\approx 0.18$, regardless the isotope content. A pronounced SST
was observed in the insulating phase (in the samples with $y >
y_{cr}$), whereas in the nearly-metallic phase (at $y < y_{cr}$),
the magnetic properties were quite different, without any
indications of a temperature-induced SST. Using the magnetic,
electrical, and thermal data, we constructed the phase diagram for
this material. The characteristic feature of this phase diagram
is a broad crossover range near $y_{cr}$ corresponding to a
competition of the phases mentioned above.  The $^{16}$O $\to
^{18}$O substitution gives rise to an increase in temperature
$T_{SS}$ of the SST and to a slight decrease in the ferromagnetic
(FM) transition temperature $T_{FM}$.

However, a number of problems important for understanding the
physics of the systems with spin-state transitions have not been
touched upon in the study reported in Ref.~\onlinecite{Kalinov2010}. The most important question is the relation between the changes caused by varying the composition (increase of concentration $y$ of the smaller rare-earth ions Eu) and that due to isotope substitution, and the physical mechanism underlying these changes. From the
phase diagram obtained in Ref.~\onlinecite{Kalinov2010} and in the present paper, we see that there exists some correlation between these changes, but the situation is not so simple: in the right part of the phase diagram, see Fig.~\ref{PhDiag} below, the SST temperature increases both with the increase of Eu content $y$ and with the increase of isotope mass (going from $^{16}$O to $^{18}$O). At the same time, in the left part of this phase diagram the effect of increasing the Eu content and of increasing the oxygen mass on the phase transition (which is then the transition to a nearly
ferromagnetic state) is just the opposite: an increase in Eu
content leads to a decrease in $T_{FM}$, but the increase of
oxygen mass -- to the increase in $T_{FM}$.

Another important open question concerns the behavior of separate
phases in the regime of phase separation. There are many different
correlated systems, in which phase separation was detected in some
range of compositions, temperatures, external fields, etc.
Typically, the measured transition temperatures in this case
changes e.g. with doping. However, it often remains unclear
whether this change is the effect occurring in separate regions of
different phases, or is just the result of averaging out over the
inhomogeneous system. To answer these questions, we now carried
out the  detailed study of the behavior of (PrEu)CoO$_3$,
using as a tool the possibility of fine tuning the properties of
the system by partial isotope substitution. This partial
substitution plays in effect the role similar to that of doping,
external pressure, etc. The obtained results  established the
possibility of ``rescaling" the changes in the system with doping
and with isotope substitution and allowed us to clarify the
questions formulated above.

As regards the second question formulated above, just the  possibility of fine tuning the properties of the system inside the region of phase separation, provided by partial isotope substitution, allows studying the behavior of different phases within this phase separated regime individually, which would be very difficult to get by other means. Our results obtained in this way demonstrate that not only the average critical temperatures  change with doping and with isotope substitution, but also  ``individual" transition temperatures (the ferromagnetic transition temperature in more metallic regions and the SST temperature in more insulating parts of the sample) do change with chemical and isotope composition.

As regards the main, the first question formulated above, about the
mechanisms governing the change of properties of the system with
chemical and isotope composition, the experimental findings
reported in the present paper provided us an opportunity to
formulate a realistic theoretical model clarifying the mechanisms
underlying the pronounced isotope effects in cobaltites exhibiting
spin-state transitions.  The theoretical analysis demonstrates
that the main factor is the change of the effective bandwidth with
the change both of chemical and isotope composition. The opposite
trends in two parts of phase diagram mentioned above find natural
explanation in this picture.

To analyze the effects of the partial oxygen isotope substitution
for the doped cobaltites in the crossover region of the phase
diagram, we have prepared a series of oxide materials with nearly
continuous tuning of their characteristics. This allows tracing the evolution of relative content of different phases as a function of the ratio $^{18}$O/$^{16}$O of the contents of oxygen isotopes. Note that there were only few investigations of this kind, one of which we undertook earlier for (La$_{1-y}$Pr$_y$)$_{0.7}$Ca$_{0.3}$MnO$_3$
manganites~\cite{Babushkina2000}. Here, the pronounced isotope
effect manifesting itself in
(Pr$_{1-y}$Eu$_y$)$_{0.7}$Ca$_{0.3}$CoO$_3$ cobaltites indeed provides us with a unique possibility to address the problems discussed above through the use of the partial oxygen isotope substitution.

\section{Experimental}

Polycrystalline (Pr$_{1-y}$Eu$_y$)$_{0.7}$Ca$_{0.3}$CoO$_3$
samples were prepared by the chemical homogenization (``paper
synthesis") method~\cite{Balagurov99} through the use of the
following operations. At first, non-concentrated water solutions
of metal nitrates Pr(NO$_3$)$_3$, Eu(NO$_3$)$_3$, Ca(NO$_3$)$_2$,
and Co(NO$_3$)$_2$ of 99.95\% purity were prepared. The exact
concentration of dissolved chemicals was established by
gravimetric titration and, in the case of Co-based solution, by
means of potentiometric titration. The weighted amounts of metal
nitrate solutions were mixed in stoichiometric ratio and the
calculated mixture of nitrates were dropped onto the ash-free
paper filters. The filters were dried out at about 80$^{\circ}$C
and the procedure of the solutions dropping was performed
repeatedly. Then, the filters were burned out and the remaining
ash was thoroughly ground. It was annealed at 800$^{\circ}$C for 2
h to remove carbon. The powder obtained was pressed into the
pellets and sintered at 1000$^{\circ}$C in the oxygen atmosphere
for 100 h. Finally, the samples were slowly cooled down to
room temperature by switching off the furnace.

Samples were analyzed at room temperature by the powder X-ray
diffraction using Cu$K_{\alpha}$ radiation. All detectable peaks
were indexed by the $Pnma$ space group. According to the X-ray
diffraction patterns, all
(Pr$_{1-y}$Eu$_y$)$_{0.7}$Ca$_{0.3}$CoO$_3$ samples were obtained
as single-phase polycrystalline materials.

We prepared a series of ceramic cobaltite samples with the degrees
of enrichment by $^{18}$O equal to 90\%, 67\%, 43\%, 17\%, and
0\%. These values were determined by the changes in the sample
mass in the course of the isotope exchange and by the mass
spectrometry of the residual gas in the oxygen exchange contour.
The samples were annealed in the appropriate $^{16}$O--$^{18}$O
gas mixture at 950~$^\circ$C during 48 h at total pressure of 1
bar. The similarity of the oxygen isotope composition in the
sample to that in the gas medium indicated that a thermodynamic
equilibrium was achieved during annealing and, hence, the
difference in the diffusion rates of the oxygen isotopes did not
significantly affect the results of the investigation. We
also note that the mass of a sample annealed in $^{16}$O$_2$
remained unchanged within the experimental error during the
prolonged heat treatment. Therefore, we can conclude that the annealing procedure does not change the oxygen stoichiometry in the
compounds under study.

The Eu doping of these samples was chosen to be near and at the
both sides of the crossover doping level $y_{cr}$: a composition
with the high Eu content exhibiting the SST, a low-Eu composition
corresponding to the nearly ferromagnetic state, and the sample at
the phase crossover, where both $T_{SS}$ and $T_{FM}$, were
observed.

Note here that the set of samples described above was specially
prepared for the present study. It turned out that the data
obtained for these samples differ somehow from that reported in
Ref.~\onlinecite{Kalinov2010} (samples of set 1). In particular, the values of the transition temperatures between phases are different for these two sets of samples. We thoroughly analyzed the possible
causes of such difference. The samples of both sets prepared by
the same technique do not differ by X-ray diffraction data, have
the same oxygen stoichiometry, but the calcium content in set 1
appeared to be lower than that corresponding to the nominal
composition. At the same time, the samples of set 2 correspond
with a high accuracy  to the chemical formula
(Pr$_{1-y}$Eu$_y$)$_{0.7}$Ca$_{0.3}$CoO$_3$. As will be discussed
below, this leads to some general shift of the phase diagram
toward larger values of of the average ionic radius $\langle r_A
\rangle$ (i.e. to the lower Eu content). Nevertheless, the general
form of the phase diagram remains the same.

For all samples, we measured the temperature dependence of the
real $\chi'(T)$ and imaginary $\chi''(T)$ parts of the ac magnetic
susceptibility and electrical resistivity $\rho(T)$. The
resistivity measurements discussed below were taken on cooling the
samples. The measurements of ac magnetic susceptibility $\chi(T)$
were performed in ac magnetic field with a frequency 667 Hz and an
amplitude of about 5 Oe in the dc magnetic field of the Earth.
Based on these measurements, we were able to analyze the evolution
of the sample properties with the change Eu and $^{18}$O content.
For achieving a better reliability, the isotope shifts $\Delta
T_{FM}=T_{FM}(M) - T_{FM}(^{16}$O) and $\Delta T_{SS}= T_{SS}(M)-
T_{SS}(^{16}$O) (where $M$ is the average atomic mass of oxygen
isotopes) were determined based on the susceptibility data
obtained both on cooling and on heating. Within the experimental
error, we did not observe the temperature hysteresis in the
vicinity of the transition to the ferromagnetic phase, whereas
near the SST, the hysteresis did not exceed 1 K. In the figures further on, we represent the $\chi(T)$ plots corresponding to the heating of the samples. The value of $T_{FM}$ transition temperatures were determined from the minimum of the logarithmic temperature derivative of susceptibility. The $T_{SS}$ transition temperatures were determined by linear approximation as demonstrated in Fig.~\ref{Fig_chi_02}. Although the critical temperatures depend on the chosen method of determining of critical temperature, the isotopic shift is nearly insensitive to the choice of determination procedure. The resistivity of the samples was measured by the conventional four-probe technique in the temperature range from 4.2 to 300 K. The value of the metal--insulator transition temperature $T_{MI}$ can be determined from the logarithmic derivative of $R(T)$ (see inset of Fig.~\ref{Fig_rho_02}).

\section{Experimental results}

\begin{figure}[ht] \begin{center}
\includegraphics*[width=0.95\columnwidth]{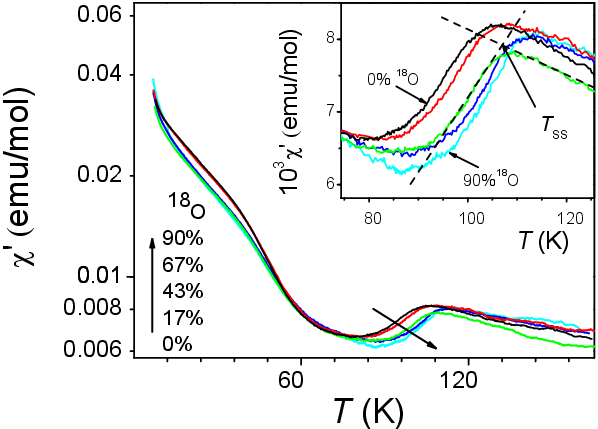} \end{center}
\caption{\label{Fig_chi_02} Temperature dependence of magnetic susceptibility for (Pr$_{1-y}$Eu$_y$)$_{0.7}$Ca$_{0.3}$CoO$_3$ with $y=0.2$. The range in the vicinity of the spin-state transition  is shown in the inset in a larger scale. The dashed straight lines illustrate determining the value of $T_{SS}$.} \end{figure}

\begin{figure} [htb] \begin{center}
\includegraphics*[width=0.95\columnwidth]{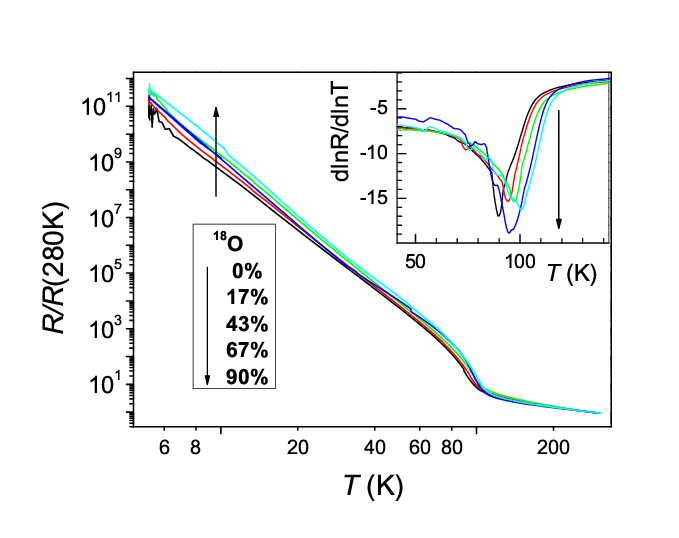} \end{center}
\caption{\label{Fig_rho_02}  Temperature dependence
of electrical resistivity for (Pr$_{1-y}$Eu$_y$)$_{0.7}$Ca$_{0.3}$CoO$_3$ with $y=0.2$. The
inset illustrates the behavior of the logarithmic derivative of
$R(T)$ used to determine the values of $T_{MI}$.} \end{figure}

{\bf 1.} For the samples with $y > y_{cr}$ (Eu content $y=0.20$), the material correspond to a ``weakly magnetic insulator", in notation
of Ref.~\onlinecite{Kalinov2010}. The $\chi'(T)$ curves give clear
indications of an SST at $T_{SS}$ manifesting itself as a peak in  $\chi'(T)$ (see Fig.~\ref{Fig_chi_02}). We see that $T_{SS}$ increases with the $^{18}$O content. The high-temperature phase is a paramagnet and a relatively good conductor, (see Fig.~\ref{Fig_rho_02}).

The low-spin (LS) insulating phase is dominant below the crossover
temperature of about 100 K.  The increase in $\chi'$ at low
temperatures is most probably caused by an incomplete transition,
after which there may remain small magnetic (and presumably more
conducting) clusters immersed into the LS state bulk insulator. As
a result of this crossover to a LS state, the electrical
resistance $R$ increases by 10--12 orders of magnitude, which can
be treated as a metal--insulator transition (MI), see
Fig.~\ref{Fig_rho_02}). In these compounds, the metal--insulator
transition is accompanied (or caused) by the SST. Note here that
the studies of the oxygen isotope effect are performed in most
cases using low-density ceramic samples with an open porosity.
This gives rise to the granularity effects related to the
existence of thin spacers at the grain boundaries  with the
properties somehow different from those of the grains themselves.
In the magnetic oxides under study, the phase separation coming
from the competition between different kinds of the exchange
interaction also manifests itself.

\begin{figure} \begin{center}
\includegraphics*[width=0.95\columnwidth]{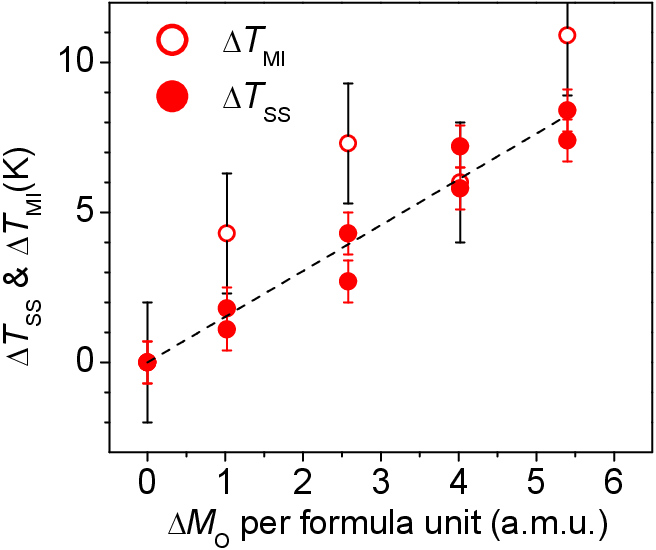} \end{center}
\caption{\label{Fig_Tss_y02} Isotopic shift of the characteristic temperatures of the spin-state $T_{SS}$ and metal--insulator $T_{MI}$ transitions at the $^{16}$O $\to ^{18}$O substitution for the sample with $y=0.2$. The values of $T_{SS}$ were determined both on cooling and on heating the samples (two closed circles at each isotope content). } \end{figure}

All these factors lead to significant changes in the temperature
dependence of electrical resistivity, so that the temperature
derivative of $R$ remains negative for the compositions under
study. Nevertheless, the characteristic features of the $R(T)$,
which are observed in the vicinity of SST and are suppressed in
the composition range corresponding to ferromagnetism,  allow  arguing that we see the manifestation of the metal--insulator
transition in  one of the phases.

The increase of $^{18}$O content does not produce a significant effect on $R(T)/R(280K)$ at low $T$, although $R/R(280K)$ is slightly larger in the samples with heavy oxygen. The mass dependence of $T_{MI}$ correlates well with the isotopic shift of $T_{SS}$ (see
Fig.~\ref{Fig_Tss_y02}). According to the calculation of isotopic
constant $T \sim M^{-\alpha}$ ($M$ is the averaged oxygen mass);
$\alpha = -d \ln T/d\ln M = -(\Delta T/\Delta M)(M/T)$, we have
the value $\alpha_{SS}$ and $\alpha_{MI} = -(0.66 \pm 0.07)$. Increasing the oxygen mass promotes the development of the LS
state.

{\bf 2.} The most important results are obtained for the samples with Eu content $y \sim y_{cr}$ ($y=0.14$ and 0.16). They correspond to a
wide concentration range of the phase separation.

\begin{figure} \begin{center}
\includegraphics*[width=0.95\columnwidth]{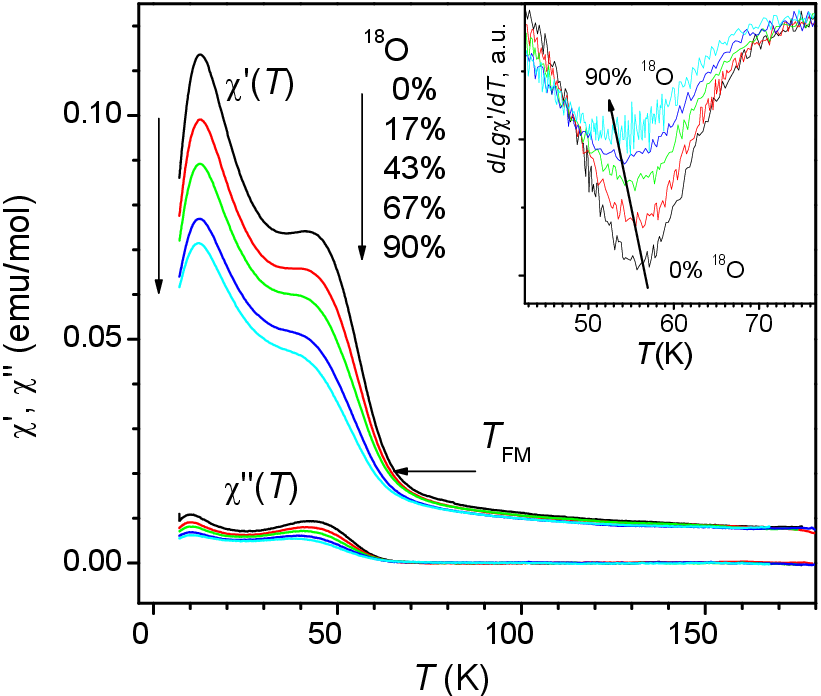} \end{center}
\caption{\label{Fig_chi_014}  Temperature dependence of magnetic susceptibility for (Pr$_{1-y}$Eu$_y$)$_{0.7}$Ca$_{0.3}$CoO$_3$ with $y=0.14$. The inset illustrates the behavior of the logarithmic derivative of $\chi'(T)$ in the vicinity of $T_{FM}$ used to determine the values of $T_{FM}$.} \end{figure}

\begin{figure} \begin{center}
\includegraphics*[width=0.95\columnwidth]{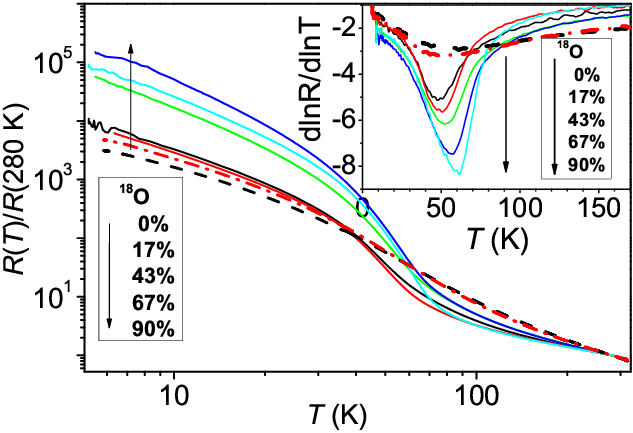} \end{center}
\caption{\label{Fig_rho_014}  Temperature dependence
of electrical resistivity for (Pr$_{1-y}$Eu$_y$)$_{0.7}$Ca$_{0.3}$CoO$_3$ with $y=0.14$ (solid
line) and $y=0.1$ with $^{16}$O (dashed line) and $^{18}$O
(dash-dotted line). The inset illustrates the behavior of the
logarithmic derivative of $R(T)$ used to determine the values of
$T_{MI}$.} \end{figure}

For the sample with $y=0.14$, we observe in $\chi'(T)$ curves a
feature  corresponding to a steep increase of the magnetization on
cooling at 60--70 K (see Fig.~\ref{Fig_chi_014}). This behavior is
caused by the FM phase arising in these samples. In addition, in
the temperature dependence of resistivity, we see a steep increase
of resistivity characteristic of the metal--insulator transition
similar to that observed for the samples with  $y = 0.2$ (see
Fig.~\ref{Fig_rho_014}). The transition temperature corresponds to
the minimum of the logarithmic derivative of the resistivity (see
the inset in Fig.~\ref{Fig_rho_014}. This means that here we deal also with the change in the relative content of metallic and insulating phases suggesting the existence of the regions corresponding to the LS insulating state and the correlation between the MI and SS transitions. At the same time we do not observe any clear indications of the SST in the temperature dependence of the magnetic susceptibility. Thus, the samples with $y =0.14$ turn out to be at the boundary of the phase separation range. The increasing in the oxygen mass favors the LS state as well as suppression of the FM phase and of the metallization.

\begin{figure} \begin{center}
\includegraphics*[width=0.95\columnwidth]{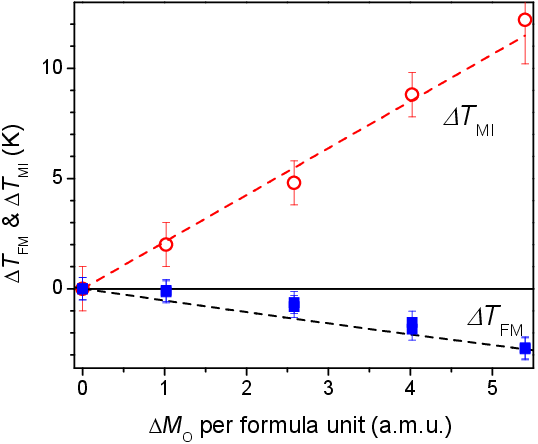} \end{center}
\caption{\label{Fig_Tss_y014}  Isotopic shift of the characteristic temperatures of the metal--insulator $T_{MI}$ (and ferromagnetic $T_{FM}$ transitions for the sample with $y=0.14$. The spin-state transition temperature $T_{SS}$ actually coincides with $T_{MI}$. The values of $T_{FM}$ were determined both on cooling and heating the samples (two nearly coinciding closed squares at each isotope content).} \end{figure}

The values of the isotopic constant calculated for the MI and FM
transitions in this sample are $\alpha_{MI} = -(2.1 \pm 0.1)$
and $\alpha_{FM} = (0.5 \pm 0.05)$, respectively. For this sample,
the isotope shifts of the characteristic transition temperatures
are illustrated in Fig.~\ref{Fig_Tss_y014}.

\begin{figure} \begin{center}
\includegraphics*[width=0.95\columnwidth]{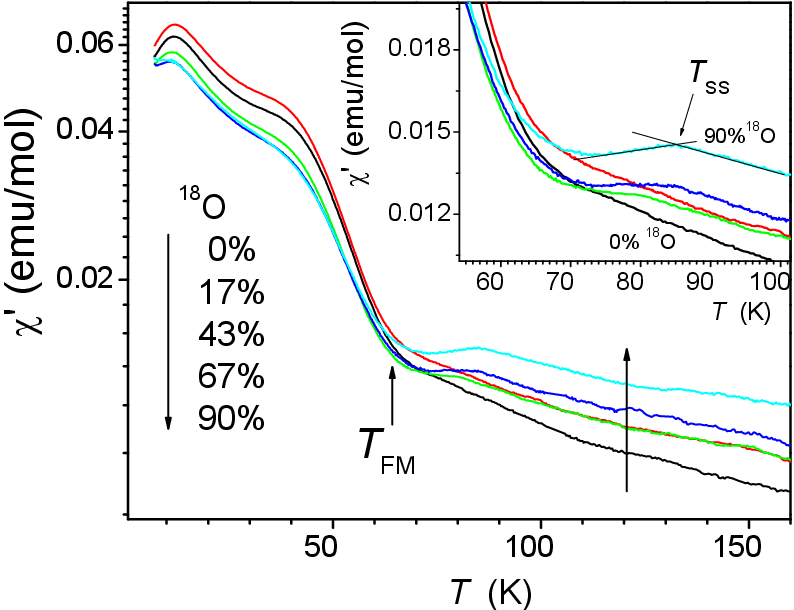} \end{center}
\caption{\label{Fig_chi_016} Temperature dependence of magnetic susceptibility for (Pr$_{1-y}$Eu$_y$)$_{0.7}$Ca$_{0.3}$CoO$_3$ with $y=0.16$. The inset illustrates the behavior of $\chi'(T)$ in a larger scale. The straight lines in the inset illustrate determining the value of $T_{SS}$.} \end{figure}

For the samples with $y =0.16$, the effect of the partial oxygen
isotope substitution by $^{18}$O manifests itself even clearer. Here, we observed the features characteristic both of FM and SS transitions. On the one hand, the temperatures dependence
of $\chi'$ exhibits a steep growth at 60--70 K similar to that in
the sample 0.14 indicating the existence of the FM transition. On
the other hand, in $\chi'(T)$, we observed a clearly pronounced
peak, which can be attributed to SST at $T_{SS}$
(Fig.~\ref{Fig_chi_016}). The $\chi'(T)$ curves also demonstrate that the transition to the LS state gradually disappears as the oxygen mass decreases. In samples with small $^{18}$O content ($<17$\%), this transition is hardly seen due to the gradual transformation from the LS to FM state when the oxygen mass decreases. The curves for samples with 17\% and 43\% of $^{18}$O corresponding to the crossover range between different phase states nearly coincide. Note also that for $y = 0.16$ (as well as in the samples with $y =0.14$) the temperature $T_{FM}$ decreases as of the average mass of oxygen increses.

\begin{figure} \begin{center}
\includegraphics*[width=0.95\columnwidth]{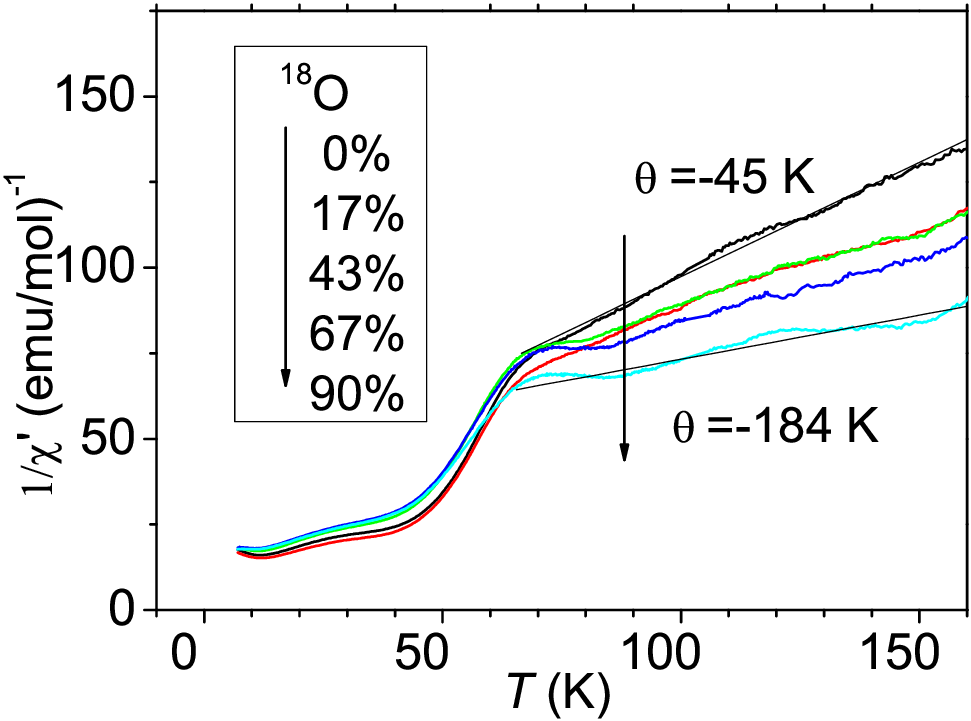} \end{center}
\caption{\label{Fig_revchi_016}  Temperature dependence of inverse magnetic susceptibility for
(Pr$_{1-y}$Eu$_y$)$_{0.7}$Ca$_{0.3}$CoO$_3$ with $y=0.16$.}
\end{figure}

In addition, the Curie--Weiss temperature $\theta$ in the formulas
for the inverse magnetic susceptibility $\chi^{-1}(T)$
considerably decreases with the increase in the average mass of
oxygen isotopes (see Fig.~\ref{Fig_revchi_016}). For the samples
with the largest oxygen mass, we have $\theta = - 184$ K, whereas
for the samples with $^{16}$O, $\theta = - 45$ K. This
phenomenon may be related to the transition from the
antiferromagnetic interaction to ferromagnetic one. Note that
Fig.~\ref{Fig_chi_014} demonstrates that $\chi''$ in the
paramagnetic range is nearly zero; this means that the absorption
related to the itinerant charge carriers is very small and does
not produce a significant effect on $\chi'$ and hence on the
values of $\theta$.

\begin{figure} \begin{center}
\includegraphics*[width=0.95\columnwidth]{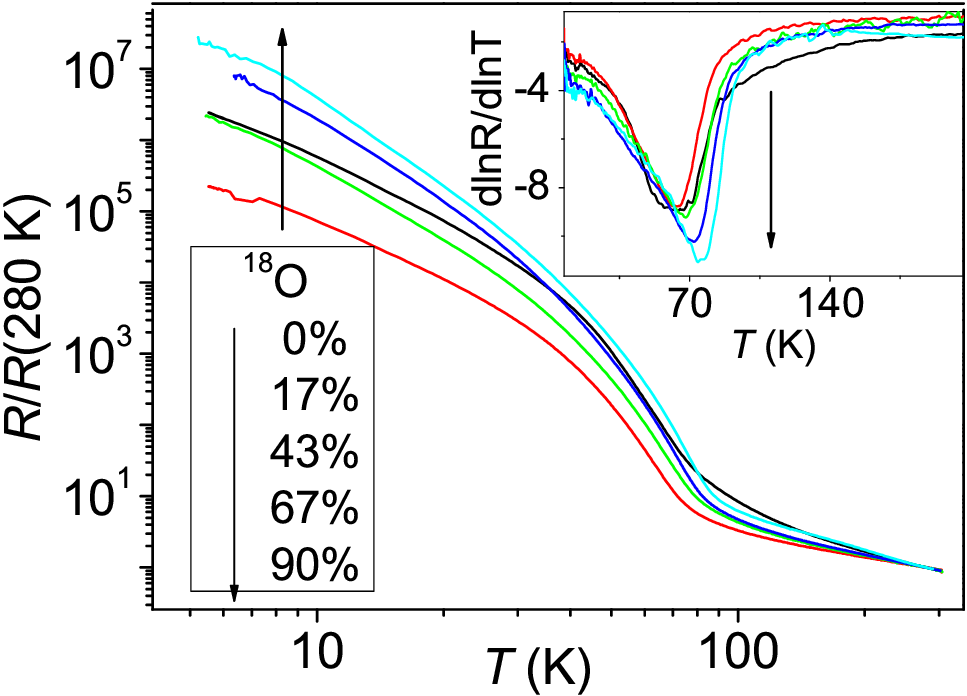} \end{center}
\caption{\label{Fig_rho_016}  Temperature dependence of electrical resistivity for (Pr$_{1-y}$Eu$_y$)$_{0.7}$Ca$_{0.3}$CoO$_3$ with $y=0.16$. The inset illustrates the behavior of the logarithmic derivative of $R(T)$ used to determine the values of $T_{MI}$.} \end{figure}

\begin{figure} \begin{center}
\includegraphics*[width=0.95\columnwidth]{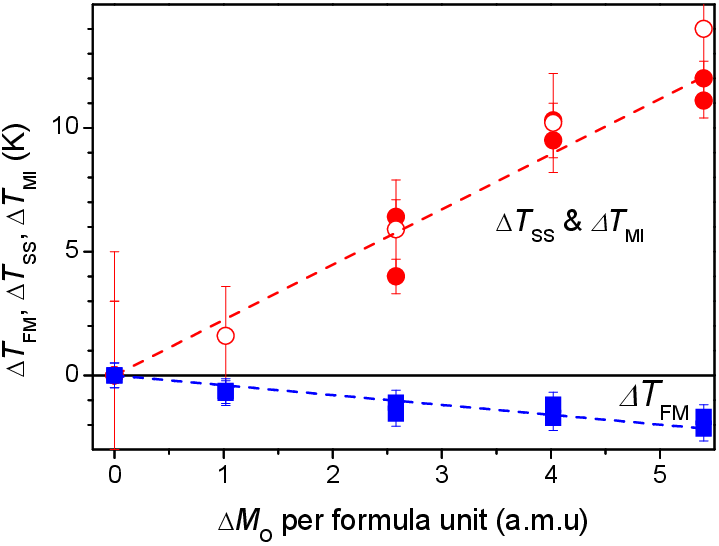} \end{center}
\caption{\label{Fig_Tss_y016}  Isotopic shift of the characteristic temperatures of the spin-state $T_{SS}$, metal--insulator $T_{MI}$ and ferromagnetic $T_{FM}$ transitions for the sample with $y=0.16$. The values of $T_{SS}$ and $T_{FM}$ were determined both on cooling and heating the samples (two dark circles or squares at each isotope content). Light circles correspond to $T_{MI}$.} \end{figure}

For the sample with $y = 0.16$ as well as for samples with $y =
0.14$ and 0.2, we have found the resistivity increases as
the temperature decreases (by 5-7 orders of magnitude), with the
MI transition in the vicinity of 70 K
(Fig.~\ref{Fig_rho_016}). Both the resistivity and magnetic
susceptibility data clearly indicate that this sample is in
the phase-separation range. The temperatures of MI and SS
transitions increase with the average mass oxygen (see the inset in
Fig.~\ref{Fig_rho_016}). In the sample with $y = 0.16$, we see the
same general tendency, namely the increase in $T_{SS}$ and the
decrease of $T_{FM}$ as the average oxygen isotope mass increases. Here, the values of the  isotope constant are
$\alpha_{SS,MI} = -(1.7 \pm 0.06)$ and $\alpha_{FM} = (0.34 \pm
0.1)$; the isotopic shifts of the characteristic transition temperatures are illustrated in Fig.~\ref{Fig_Tss_y016}.

{\bf 3.} Finally, the samples with $y < y_{cr}$ (Eu content $y=0.10$) fall into the range of ``nearly metallic" ferromagnet. In Ref.~\onlinecite{Kalinov2010}, it was shown that at $T < T_{FM}$ the
compositions with a low Eu content  correspond to the domains of
the metallic ferromagnetic phase embedded in a weakly magnetic
nonconducting matrix.

\begin{figure} \begin{center}
\includegraphics*[width=0.95\columnwidth]{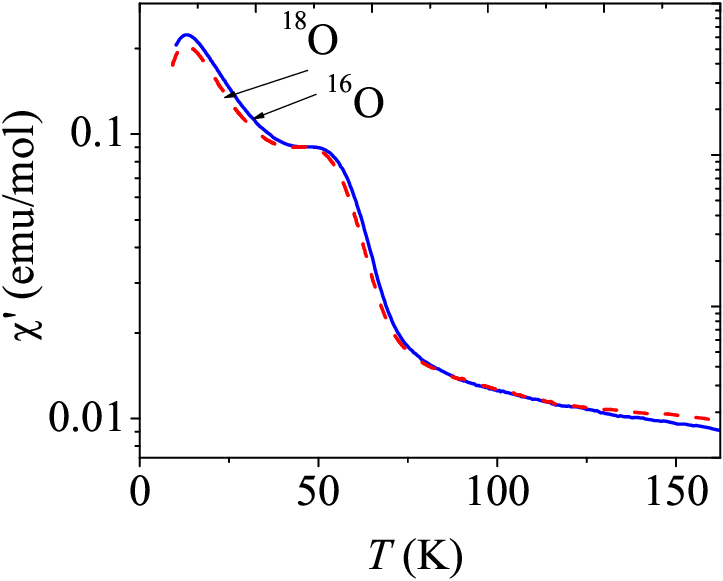} \end{center}
\caption{\label{Fig_chi_01} Temperature dependence of magnetic susceptibility for (Pr$_{1-y}$Eu$_y$)$_{0.7}$Ca$_{0.3}$CoO$_3$ with $y=0.1$.}
\end{figure}

According to the temperature dependence $\chi'(T)$ plotted in
Fig.~\ref{Fig_chi_01}, the magnetization steeply increases on
cooling at about $60-70$ K (i.e. the ferromagnetic phase arises).
With the increase in the average oxygen mass,  the value of
$\chi'(T)$ decreases at low temperatures. Here, we have
$T_{FM}$($^{18}$O) $< T_{FM}$($^{16}$O) and the maximum isotope
shift of $T_{FM}$ does not exceed 2--3 K.

The electrical resistivity for these samples ($y=0.10$) increases
at low temperatures, even when the ferromagnetic phase arises.
This behavior is quite similar to that observed in the samples
with $y=0.14$ and $y=0.16$  (see Fig.~\ref{Fig_rho_014}). We note
that the electrical resistance in the samples with $^{18}$O is
higher than in those with $^{16}$O. However, with the decrease of
Eu content down to $y =0.10$, the MI peculiarity in the resistance
is suppressed. Such a behavior of $R(T)$ for samples with $y =0.10$ can be compared with that for $y =0.14$ samples (see
Fig.~\ref{Fig_rho_014} and the inset of this figure). The resistivity for samples $y=0.10$ corresponds to a more smooth curve than for
samples with $y=0.14$, although the regular course of the temperature dependence remains nearly unchanged.  Thus, the samples with $y=0.10$ do not become truly metallic but their behavior differs from the behavior of the samples with $y=0.14$ (they do not exhibit indications of an SST). Therefore, we argue that in the phase diagram, the composition with $y=0.10$ lies outside the crossover region, on the left-hand side of it. The value of isotope constant is  $\alpha_{FM} = (0.23 \pm 0.1)$.

We note here that for such a complicated system, there is no genuinely accurate method for determining thermodynamic values of
transition temperatures $T_{SS}$  and especially $T_{FM}$.
Probably, the closest to the actual value is the onset
temperature of the transition ($T^*$). The value of $T^*$ can be
determined by different methods, but as we have already mentioned in Sec. II, the isotope shift of the transition temperature is nearly insensitive to the definition of $T^*$. Therefore, in Figs.~\ref{Fig_Tss_y02}, \ref{Fig_Tss_y014}, and
\ref{Fig_Tss_y016}, we show only the isotope shifts of the
transition temperatures and not the temperatures themselves.

\section{Discussion of experimental results}

\begin{figure}[ht] \begin{center}
\includegraphics*[width=0.95\columnwidth]{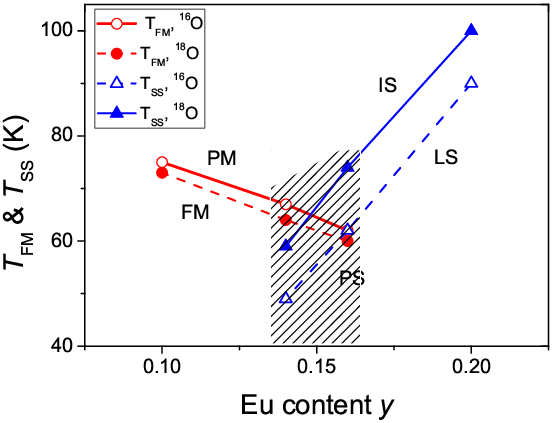} \end{center}
\caption{\label{PhDiag} Phase diagram of (Pr$_{1-y}$Eu$_y$)$_{0.7}$Ca$_{0.3}$CoO$_3$ compound with $^{16}$O
and 90\% $^{18}$O. PM, FM, IS, and LS stand for the paramagnetic,
ferromagnetic, intermediate-spin, and low-spin state,
respectively. The hatched area corresponds to the phase-separated
state (PS).} \end{figure}

The obtained data for the temperatures of the phase transitions
can be presented in the form of a phase diagram (Fig.~\ref{PhDiag}). It illustrates that as the of Eu content increases, the system transforms from the nearly metallic ferromagnet to the LS insulating state undergoing LS $\to$ IS spin-state transitions~\cite{ISstate}. Between these states, we have a broad crossover region corresponding to the phase separation. Indeed, the simultaneous observation both of $T_{FM}$ and $T_{SS}$ in the samples with Eu content $y =0.14$ and 0.16 is a clear evidence of the phase separation in the system.

These samples also provide a spectacular illustration of the
effect related to the variable content of $^{18}$O. In particular,
for the samples with $y =0.16$, the temperature dependence of
magnetic susceptibility (Fig.~\ref{Fig_chi_014}) exhibits  a pronounced feature corresponding to the SST at high values of $^{18}$O content. With decreasing the $^{18}$O content, this feature
becomes weaker and disappears below 17\% of $^{18}$O. Hence, we
see that the change on the average oxygen mass can drastically
affect the phase composition of the cobaltite samples.

The oxygen isotope substitution $^{16}$O $\to ^{18}$O shifts the
phase equilibrium toward the insulating state. For the heavier
isotope, the SST temperature $T_{SS}$ increases, while the ferromagnetic transition temperature $T_{FM}$ decreases. Varying the average oxygen mass is a unique tool for investigating special properties of phase separation in cobaltites near the crossover between the FM and LS phases. We also see that the effect of increasing the $^{18}$O content on the system is similar to that of increasing the Eu content.

We emphasize that the general structure of the phase diagram
shown in Fig.~\ref{PhDiag} is similar to that reported in Ref.~\onlinecite{Kalinov2010}. The partial oxygen isotope substitution allows a much more detailed study of the evolution of the state of
doped perovskite cobaltites in the most interesting region
corresponding to the phase separation. This is exactly the main
objective of our work, which could not be significantly affected
by the difference in the composition of two sets of samples.

The analysis of these results for different chemical and isotope
composition demonstrates that the effect of the increase in the Eu
content $y$ and of the average oxygen mass are qualitatively similar. We can rescale the dependence of the transition temperature on
both parameters using the combined variable $y + 0.015x$,  for
$T_{SS}$ or $y + 0.01x$ for $T_{FM}$, where $x$ is the relative
content of $^{18}$O, as shown in Fig.~\ref{Tss(x,y)}. Both the
temperatures of SS and FM transitions depend almost linearly on
this combined variable. We see that the change of Eu content by
1\% is equivalent to the change of the isotope content by 70--100
\%. This actually the most important result of the present study.
The theoretical analysis of these results is given in the
next sections.

\begin{figure} \centering
   \subfigure[]{
      \includegraphics*[width=0.95\columnwidth]{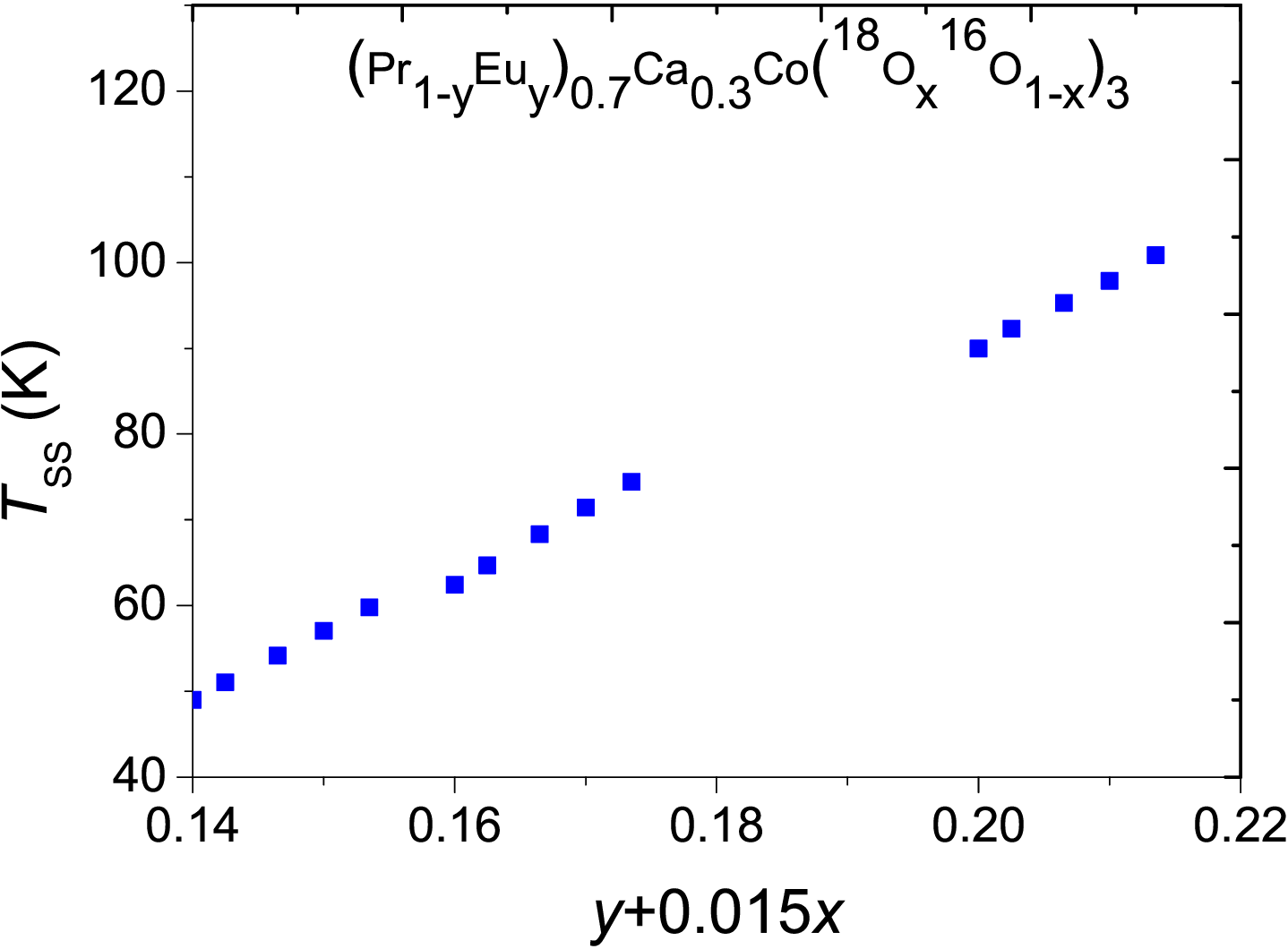}}
   \subfigure[]{
      \includegraphics*[width=0.95\columnwidth]{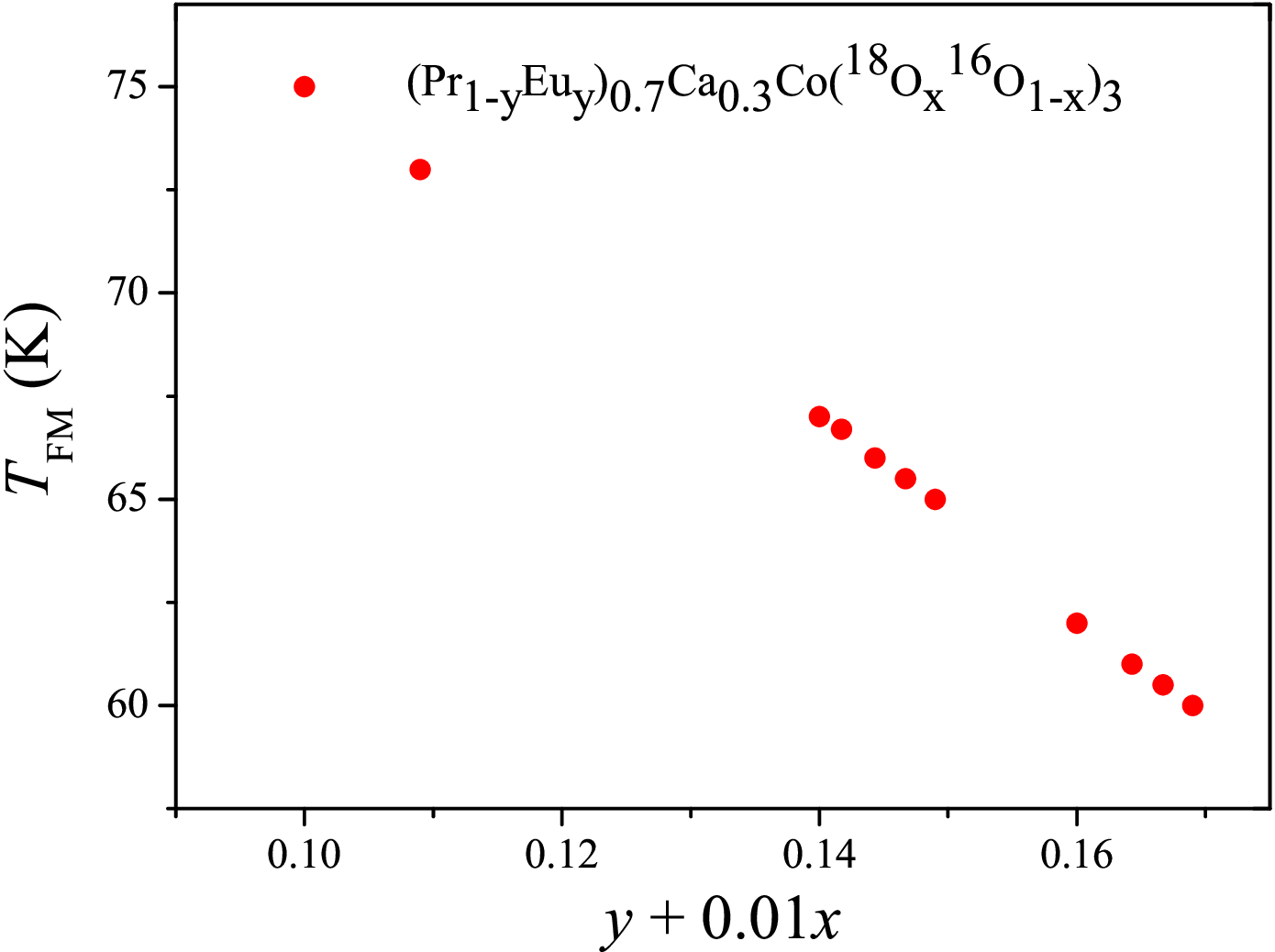}}
   \caption{ Plots illustrating the combined effect of the Eu and $^{18}$O doping on the temperature of spin-state transition $T_{SS}$ and on the ferromagnetic transition temperature $T_{FM}$. Note the difference of the temperature scales in panels (a) and (b).} \label{Tss(x,y)}
\end{figure}

\section{Calculation details}

To explain the composition and isotope dependence of the
properties of our system (see Fig.~\ref{PhDiag}), especially the
similar dependence of the SST temperature and the temperature of
the FM transition illustrated in Fig.~\ref{Tss(x,y)}, we propose a
realistic model (Section~\ref{SecIsotope})  based predominantly on
the change of the electron bandwidth with chemical and isotope
composition.  Some input, as well as the estimates of relevant
parameters are taken from the {\it ab initio} band structure
calculations for the limiting ``pure" compositions corresponding
to $y =0$ (PrCoO$_3$) and $y =1$ (EuCoO$_3$).

The crystal structure of PrCoO$_3$ obtained in Ref.~\onlinecite{Knizek2009} for $T=300$~K was utilized in those calculations. For EuCoO$_3$, the lattice parameters were taken from Ref.~\onlinecite{Baier2005}. The exact atomic positions for EuCoO$_3$ are unknown, and we therefore used the same positions as for PrCoO$_3$ (with the correct unit cell volume for EuCoO$_3$). The splitting between different one-electron energy levels $\Delta_{CF}$ was calculated within the local density approximation (LDA) in the framework of the method of linear muffin-tin orbitals  (LMTO)~\cite{Andersen1984}. Partially filled but physically unimportant $4f$ states of the Eu and Pr were treated as frozen~\cite{Nekrasov2003}.

The Brillouin-zone (BZ) integration in the course of
self-consistency iterations was performed over a mesh of 144 {\bf
k}-points in the irreducible part of the BZ.

\section{Doping dependence: LDA results}

There are different ways to estimate the SST temperature  with the use of the band structure calculation. The most direct way is to calculate the total energies of different spin states~\cite{Korotin96,Streltsov11}. However, in the case of the doped system this would require very large supercells. Moreover, currently, we have no single commonly accepted model that can explain all experimental facts. Various combinations of a static or dynamic order of the different spin states are discussed in the
literature~\cite{Nekrasov2003,Knizek2009-2,Haverkort-06,Ren-11,Lamonova-11}.
This is the reason why we chose an alternative approach.

\begin{figure}
 \centering
 \includegraphics*[width=0.95\columnwidth]{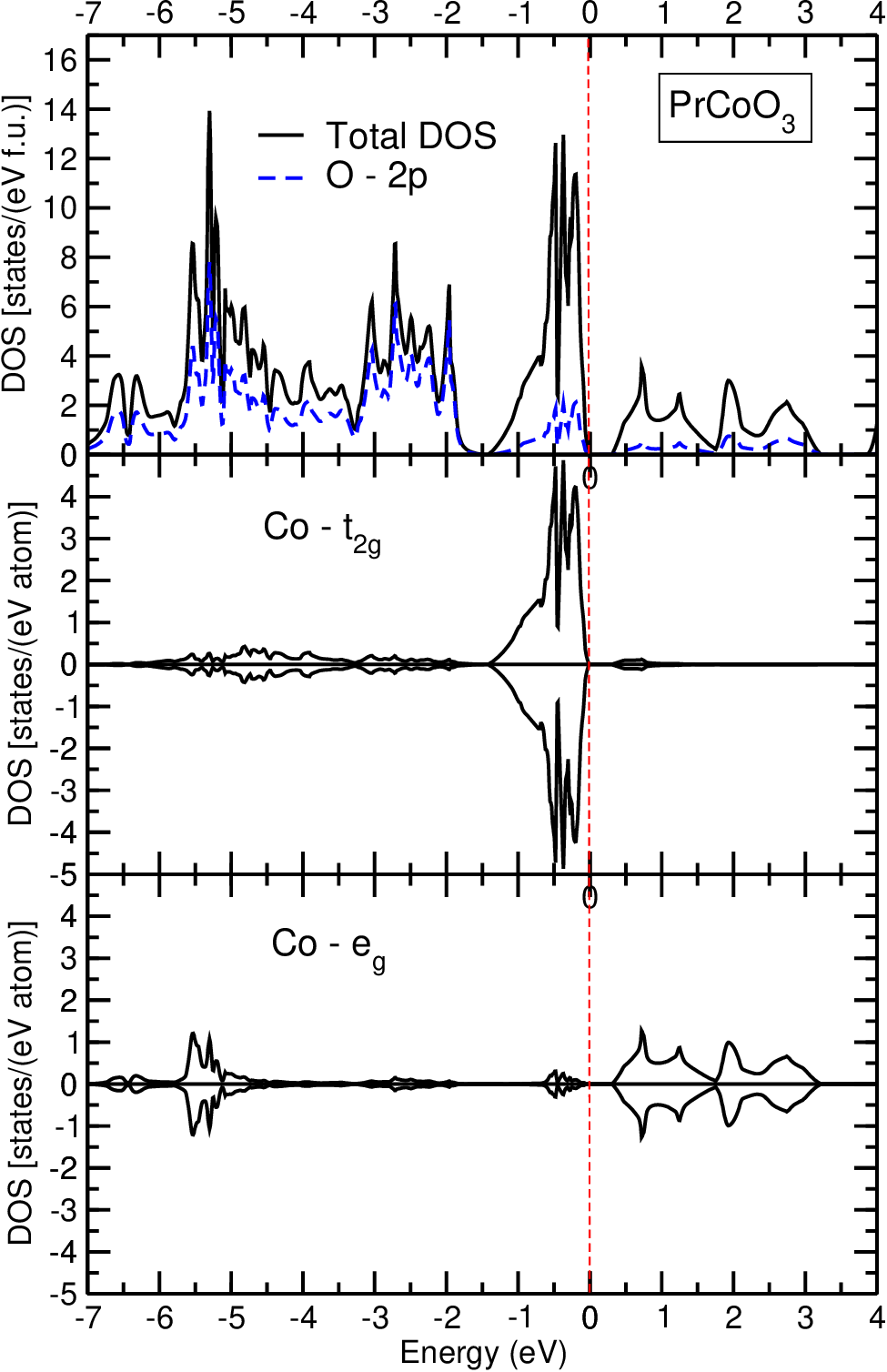}
\caption{\label{PrCoO3} The total  and partial densities of states (DOS) for PrCoO$_3$. The Fermi energy corresponds to zero.} \end{figure}

The energy of any of the spin states depends on two important
parameters: the single-electron energy difference $\Delta_{CFS}$
between the highest $t_{2g}$ and the lowest $e_g$ levels, and the
intra-atomic Hund's rule exchange coupling $J_H$. The Hund's rule
energy $J_H$ is an atomic characteristic and does not change
appreciably either with the Eu doping or with the isotope
substitution. Hence, to investigate the dependence of the SST temperature on the doping or isotope substitution in the first approximation, we can focus on the study of the single
parameter, $\Delta_{CFS}$. If $\Delta_{CFS}$ is large enough
($\Delta_{CFS} > 2J_H$), it is energetically favorable to
localize all electrons in the low-lying $t_{2g}$ subshell of
Co$^{3+}$, i.e., the system is in the LS state. With a decrease of the crystal-field splitting, some electrons can be transferred to the $e_g$ subshell, which allows the system to gain the exchange energy, since there are more electrons with the same spin. As we will show below, both the isotope substitution and the doping can be related to the crystal-field splitting.

In order to estimate $\Delta_{CFS}$, we used the Wannier function
projection procedure proposed in Ref.~\onlinecite{Streltsov2005}, which allows projecting the full-orbital band Hamiltonian onto the
subspace of a few states (five $d$ states of Co). With the Fourier
transformation, we obtain the Hamiltonian in real space, from
which the splitting between highest in energy $t_{2g}$ and lowest
$e_g$ can be easily calculated. For PrCoO$_3$, we obtain
$\Delta_{CFS} = 2.07$~eV.

The total and partial densities of states (DOS) obtained for
PrCoO$_3$ in the LDA calculations are presented in
Fig.~\ref{PrCoO3}. In the octahedral symmetry, the 3$d$ states of
Co are split into $t_{2g}$ and $e_g$ subbands. In the LDA, the
valence band is mostly formed by the Co$-t_{2g}$ states, while the
conduction band is determined by the Co$-e_{g}$ states. The O$-2p$
band is located in the energy range from --7 to --1.5~eV.

The DOS for EuCoO$_3$ is qualitatively very similar and
corresponding calculations of $\Delta_{CFS}$ result in the value
of 2.14~eV. The increase of the $t_{2g}-e_g$ excitation energy on
going from PrCoO$_3$ to EuCoO$_3$ is caused by two factors. The
first is the lanthanide contraction: the substitution of the large
Pr$^{3+}$ by smaller Eu$^{3+}$ ions leads to some decrease of the
Co--O distance and to the corresponding increase in the $p-d$
hybridization, which leads to an increase in the difference
between the centers of the $t_{2g}$ and $e_g$ bands. The second
effect is related to the decrease in the effective widths of
$t_{2g}$ and $e_g$ energy bands with the corresponding increase in
the energy gap between them. This narrowing of energy bands on
going from PrCoO$_3$ to EuCoO$_3$ is also related, in effect, with
the lanthanide contraction, because of which the tilting of
CoO$_6$ octahedra increases and the Co--O--Co angle and the
corresponding bandwidth decrease in going from PrCoO$_3$ to
EuCoO$_3$. Both these effects eventually lead to the increase in
$\Delta_{CFS}$ with the Eu content, which leads to the enhanced
stabilization of the LS state of Co$^{3+}$ (see the more detailed
discussion of these effects in Sec.~\ref{SecIsotope}).

This change of the crystal-field splitting (CFS) results in the
modification of the SST temperature, since this transition is due to the competition of the Hund's rule exchange coupling $J_H$ and the CFS~\cite{Nekrasov2003}.

It was found in Refs.~\onlinecite{Kalinov2010} and \onlinecite{Babushkina2010} (see also Fig.~\ref{PhDiag}) that the change of the Eu content $y$ in (Pr$_{1-y}$Eu$_y$)$_{0.7}$Ca$_{0.3}$CoO$_3$ by 0.02 leads to the
change of the SST temperature by about 14 K. In the first approximation, it is possible to neglect the presence of
Ca and interpolate the change of the CFS for the complex system
like (Pr$_{1-y}$Eu$_y$)$_{0.7}$Ca$_{0.3}$CoO$_3$ using the values
of the CFS for $y=0$ (PrCoO$_3$) and $y=1$ (EuCoO$_3$). Indeed,
the substitution of Ca for Eu$^{3+}$/Pr$^{3+}$ gives rise to
ligand holes, which manifest themselves mainly in the rigid shift
of the O-2$p$ band upwards resulting in a slight increase of the
CFS. Such a linear interpolation predicts the change of the SST temperature by 16~K, if $y$ changes by 0.02, which is in an excellent agreement with the
experiment.

\section{Isotope substitution: Model results \label{SecIsotope}}

The isotope substitution does not change the chemical properties
of the ions such as the oxidation numbers or bonding energies.
However, it affects the crystal lattice through the modification
of the phonon spectra. Below, following the approach in Ref.~\onlinecite{Babushkina98}, we demonstrate the effect of this
modification on the electronic and magnetic properties and hence
on the SST.

In the tight-binding model, the band spectra of a solid is
determined by the on-site ionic energy levels
$\varepsilon_i^{nlm}$ and the hopping matrix elements between
different sites $t_{ij}^{ll',mm'}$. The ionic energies
$\varepsilon_i^{nlm}$ are obviously independent of the mass of the
ions, being determined by the quantum numbers and the intra-atomic
Coulomb and exchange interactions. The hopping parameters depend
on the type of the orbitals ($s,p,d,f$), bonding type
($\pi,\sigma,\delta$), and the distance between ions, $u$.
According to the famous Harrison
parametrization~\cite{Harrison1999,Andersen1977} in the absence
lattice vibrations, the hopping integrals, e.g., between $p$ orbitals of the oxygen and transition metal $d$-orbitals equal to

\begin{equation} \label{hopdist}
t_{pd} = \frac{C_{pdm}}{u^{4}},
\end{equation}
where coefficients $C_{pdm}$ depend on the type of the bonding and can be different for different metals and ligands~\cite{Harrison1999,Slatter-Koster}.

The static version of \eqref{hopdist} can be  generalized taking
into account the presence of lattice vibrations, i.e. phonons,
which depend on the ion masses. The mean $pd$ hopping matrix
element can be calculated as
\begin{eqnarray}
\langle t_{pd}\rangle &=& \frac 1{2v}
\int_{u_0-v}^{u_0+v} \frac{C_{pdm}}{u^4} du =\nonumber \\
&=& \frac{C_{pdm}} {6v} \Big(\frac 1{(u_0-v)^3} - \frac
1{(u_0+v)^3} \Big), \end{eqnarray}
where $v=\sqrt {\langle \delta u^2 \rangle}$ is the mean square displacement from the equilibrium position $u_0$ due to phonons. Since $v/u_0 \ll 1$, we can simplify the last equation expanding it to series to the 4th order
\begin{eqnarray} \label{hopdist-phonon} \langle t_{pd} \rangle =
\frac{C_{pdm}}{u_0^4} \Big( 1 + \frac{10}3 \Big(\frac v
{u_0}\Big)^2\Big) + O(\Big(\frac v{u_0}\Big)^4). \end{eqnarray}
In the static limit $v \to 0$, the last formula coincides with
\eqref{hopdist}.

\begin{figure} \centering
   \subfigure[]{
      \includegraphics*[width=0.95\columnwidth]{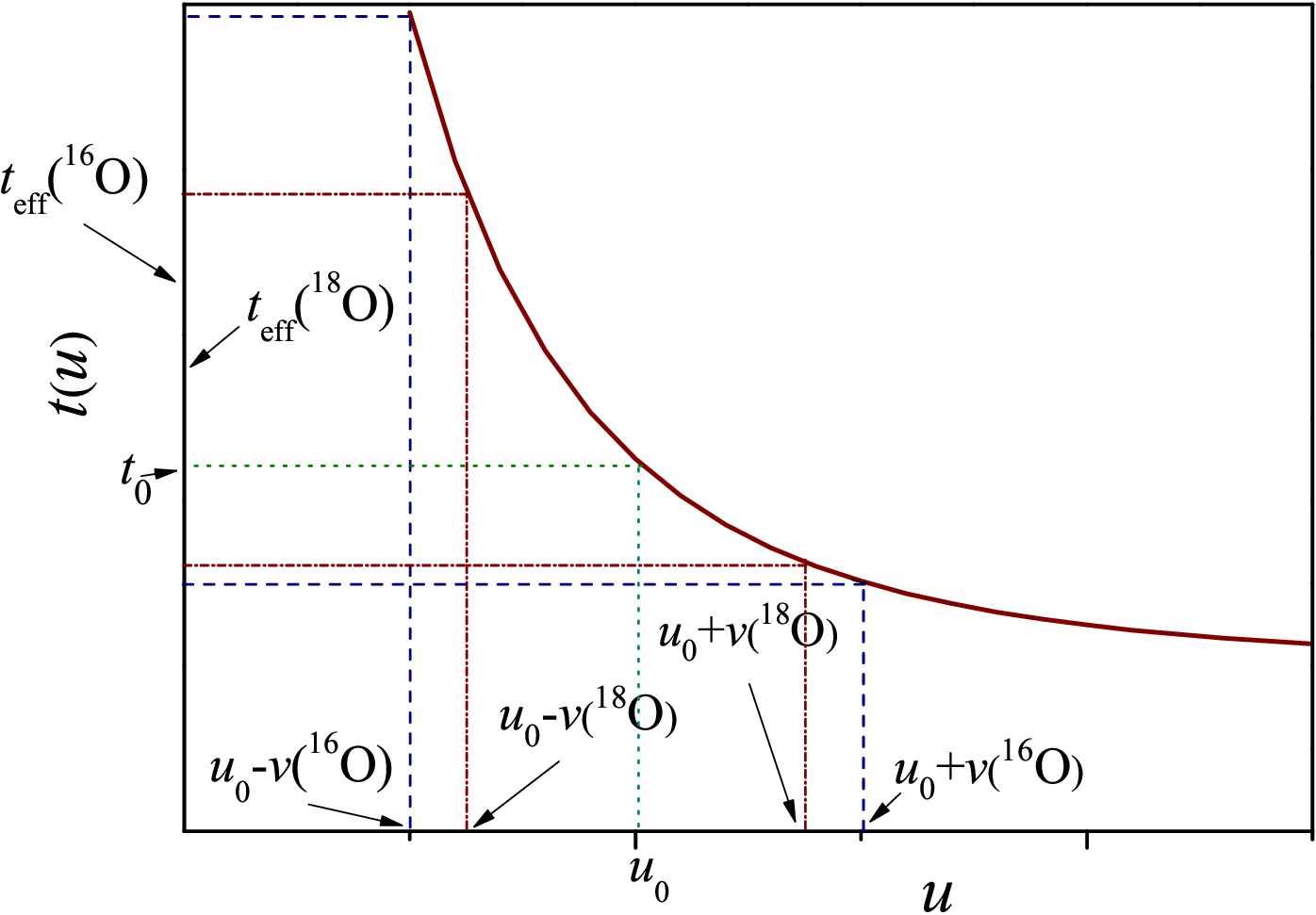}}
   \subfigure[]{
      \includegraphics*[width=0.95\columnwidth]{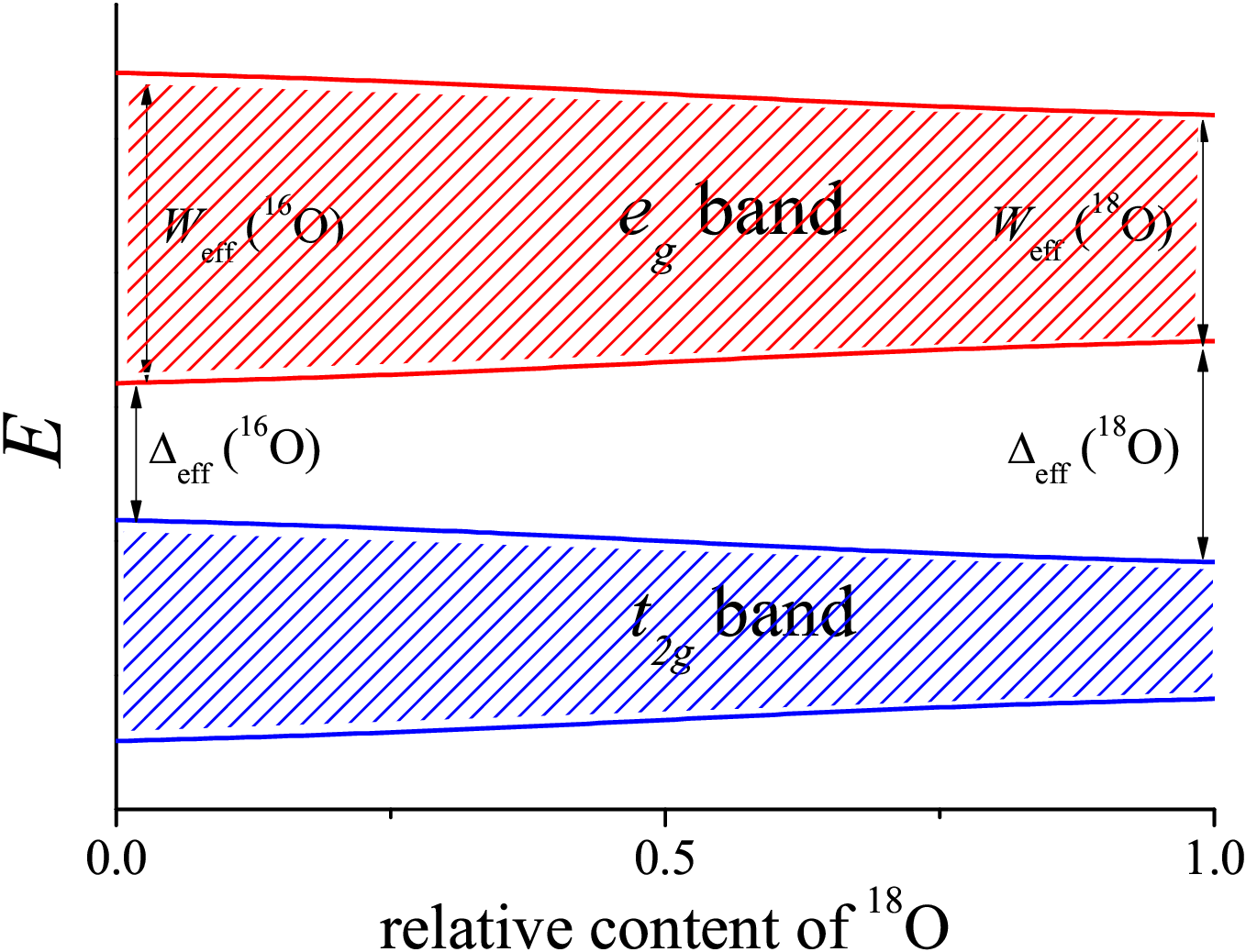}}
\caption{ Schematic illustration of the effects of oxygen isotope substitution (a) on  the effective hopping integral $t_{eff}$ and  (b) on the effective gap $\Delta_{eff}$ between the $e_g$ and $t_{2g}$ bands and on the effective width $W_{eff}$ of the $e_g$ band. The solid curve in panel (a) depicts the hopping integral $t$ as function of inter-ion distance $u$ according to \eqref{hopdist}. Effective hopping integrals $t_{eff}$ are determined by averaging $t(u)$ over the interval given by the mean square displacement $v$ of ions due to the lattice vibrations \eqref{v} (crudely, $t_{eff} = 1/2[t(u_0+v) + t(u_0-v)]$ for each isotope). We see that $t_{eff}$($^{16}$O) $> t_{eff}$($^{18}$O). As a result, $\Delta_{eff}$($^{18}$O) $> \Delta_{eff}$($^{16}$O)
and hence $T_{SS}$($^{18}$O) $> T_{SS}$($^{16}$O). At the same
time, $W_{eff}$($^{18}$O) $< W_{eff}$($^{16}$O) and hence
$T_{FM}$($^{18}$O) $< T_{FM}$($^{18}$O).} \label{Fig_model}
\end{figure}

In the Debye model at zero temperature, the mean square
displacement is written as~\cite{Reissland1973}
\begin{equation}
\label{v} v^2 = \langle \delta u^2 \rangle = \frac{9\hbar^2}{4k_B
\theta_D} \frac1{m}. \end{equation}
Here $m$ is the mass of vibrating ions and $\theta_D$ is the Debye temperature. Due to different mass the mean square displacement in the compounds enriched by $^{16}$O or $^{18}$O will be different. The Debye temperature for the very similar system, LaCoO$_3$ was found to be $\sim$600~K \cite{Stolen1997}. One may use this value to estimate
$v$ in (Pr$_{1-y}$Eu$_y$)$_{0.7}$Ca$_{0.3}$CoO$_3$. Then, the mean-square displacement is $v_{18} = 0.100$ $\AA$ for $^{18}$O and
$v_{16} = 0.107$ $\AA$ for $^{16}$O.  According to
\eqref{hopdist-phonon}, this leads to the decrease in the effective bandwidth in going from $^{16}$O to $^{18}$O. A qualitative explanation of this effect is presented in Fig.~\ref{Fig_model}a. For strong electron-phonon coupling, the same effect could be attributed to the polaron band narrowing depending on the isotope mass.

The SST temperature depends on the energy difference between $t_{2g}$ and $e_g$ subbands, which is defined by the widths of corresponding bands and the positions of their centers. We start with the study of the bandwidths dependence on the ligand ion mass (see the schematic  illustration in Fig.~\ref{Fig_model}b).

To calculate the change in the bandwidth caused by the
$^{16}$O $\to$ $^{18}$O substitution, we needs to know the
hopping integrals, which depend on two unknown parameters
$C_{pdm}$ and $u_0$.  The $C_{pdm}$ coefficients can in
principle be evaluated as it is prescribed, for instance, in
Ref.~\onlinecite{Harrison1999}. However, for a better precision, we
calculated $C_{pdm}$ parameters from the LDA $t_{2g}$ and $e_g$
bandwidths in pure PrCoO$_3$. The equilibrium Co--O distance $u_0$
in its turn can be evaluated from the actual crystal structure
data for EuCoO$_3$ and PrCoO$_3$.

Finally, performing all these calculations, one gets that the
$t_{2g}$ bandwidth decreases by 22~K when $^{16}$O is replaced by
$^{18}$O. The decrease in the $e_g$ bandwidth is two times larger
and equals 44~K. In effect, the minimum energy of the $t_{2g} \to
e_g$ transition increases by 33~K. Hence, the SST temperature due to the change of the bandwidth must increase similarly in passing from $^{16}$O to $^{18}$O in qualitative accordance with the  experiment (we even overestimate the actually observed changes; see Fig.~\ref{PhDiag}).

We consider the second mechanism, which affects the SST and is related to the dependence of changes in the centers of gravity of corresponding bands (i.e., the CFS) with the isotope substitution. It turns out that this effect counteracts the first one (change of the effective $t_{2g}$ and $e_g$ bandwidths), but this second effect is much smaller numerically (see below). Generally speaking, there are two main contributions to the CFS, $\Delta_{CFS}$. One comes from the Coulomb interaction of the $3d$ electrons with the negatively charged ligands, another is due to the hybridization between $d$ orbitals of metal ions and $p$ orbitals of the ligands~\cite{Ballhausen1962,Sugano1970}. For most oxides of $3d$
transition metals,  both terms act in ``the same direction'', resulting to the same sequence of levels~\cite{Ushakov2011}. That is why so crude approaches as the atomic sphere approximation (ASA)~\cite{Skiver1984} often used in the {\it ab initio} calculations provide quite precise band structure in most cases. The effect of the Coulomb term can be omitted or effectively incorporated into the kinetic energy contribution. Below, we follow the same strategy by considering the kinetic energy only, keeping in the mind that the Coulomb contribution can be taken into account via the parameter renormalization.

In the second order of the perturbation theory, the CFS between $t_{2g}$ and $e_g$ subbands is written as
\begin{equation} \label{deltaCF} \Delta_{CFS} =
\frac{t^2_{pd\sigma} - t^2_{pd\pi}}{\Delta_{CT}}, \end{equation}
where $\Delta_{CT}$ is the charge-transfer energy (which
corresponds to the $d^n p^6 \to d^{n+1} p^5$ transition),
$t_{pd\sigma}$ and $t_{pd\pi}$ are the hopping matrix elements for
different types of bonds.

Because the average hopping $\langle t_{pd} \rangle$ decreases in
passing from $^{16}$O to $^{18}$O according to Eqs.~\eqref{hopdist-phonon} and~\eqref{v}, the CFC should also decrease as follows from Eq.~\eqref{deltaCF}. As a result, this contribution should lead to the opposite tendency: a decrease of the SST transition temperature in going from $^{16}$O to $^{18}$O. However, this effect does not exceed a few kelvins at realistic values of the charge-transfer energy in cobaltites~\cite{Chainani1992}.

Note also that we estimated here the changes in the distance
between the {\it edges} of $t_{2g}$ and $e_g$ subbands. However,
at finite temperatures, we must have not only the transitions
between the band edges but just from one subband to another. Such a
temperature-induced smearing could diminish somehow our estimates
of the isotope effect in SST.

In Fig.~\ref{Tss(x,y)}b, we see that the isotope effect for
$T_{FM}$, being much weaker than for $T_{SS}$, is of the
opposite sign. Nevertheless, there is the same similarity between
the effects of the Eu content and the oxygen isotope substitution.
This is in agreement with our expectations, because the
ferromagnetism of the low-Eu doped samples with metallic clusters
should be stabilized by the double-exchange mechanism, according
to which $T_{FM}$ is proportional to the effective bandwidth of
itinerant electrons. This bandwidth decreases for the heavier
isotope, and that is why $T_{FM}$ decreases at the $^{16}$O $\to
^{18}$O substitution as well as at increasing Eu content, see
Fig.~\ref{Tss(x,y)}b. A schematic illustration of the mechanism
underlying the oxygen isotope effect discussed above is given in
Fig.~\ref{Fig_model}.

\section{Conclusions}

Experimental studies carried out for
(Pr$_{1-y}$Eu$_y$)$_{0.7}$Ca$_{0.3}$CoO$_3$ cobaltites with
varying isotope substitution of $^{16}$O by $^{18}$O demonstrated
that there exists a strong similarity in the changes caused by the
chemical composition (increasing the Eu content) and those arising
from the oxygen isotope substitution. The chemical composition $y
\sim 0.1-0.2$ was chosen because in this range a crossover occurs
between the ferromagnetic near-metallic state with magnetic Co
ions to the nonmagnetic insulator with the low-spin Co$^{3+}$
($t_{2g}^6e_g^0$, $S = 0$), see Fig.~\ref{PhDiag}.

The main experimental conclusion presented in Fig.~\ref{Tss(x,y)}
is that one can rescale the behavior of this system. The
dependence of the spin-state transition temperature and of the
ferromagnetic transition temperature on the Eu content $y$ and on
the content $x$ of the heavier isotope $^{18}$O can be represented
by the same almost linear plot as function of the combined
variables $y + 0.015x$ and $y + 0.01x$, respectively. This means,
for example, that for the the SST, the change of the Eu content $y$ by 0.007 is equivalent to the substitution of 50\% of
$^{16}$O by $^{18}$O. In addition, this clearly demonstrates that
not only the average transition temperatures change with doping
and with isotope substitution, but also the transition temperatures for each separate phase vary with chemical and isotope composition.

Based on this similarity between the role of chemical and isotope
composition for the SST and for the transition to the ferromagnetic state at a smaller Eu content, we propose a theoretical explanation of the isotope effect in these transitions. We investigate the corresponding changes and estimate the relevant parameters using the {\it ab initio} band structure calculations. These results together with the analytical model allow explaining the observed behavior. In particular, the isotope effect both in the spin-state and ferromagnetic transitions is interpreted in terms of the change in the corresponding widths of the $d$ bands occurring due to the
electron-phonon renormalization, which depends on the atomic
masses of the respective isotopes.

All this demonstrates once again that the oxygen isotope
substitution is a powerful tool for revealing salient features in
the behavior of strongly correlated magnetic oxides.

Summarizing, we can say that using this  approach, we established,
first, that in the aforementioned crossover range, we can clearly
distinguish two coexisting phases, nearly insulating exhibiting a
spin-state transition and nearly metallic ``ferromagnetic", with
the different behavior of the transition temperatures. Second, we
have found that these transition temperatures depend almost
linearly on the content of the heavy oxygen isotope, which is a
nontrivial observation clearly demonstrating that the electronic
structure could be effectively controlled by isotopes. Third,
based on these observations and using the parameters deduced from
our band-structure calculations, we put forward a simplified model
capturing the main physics of the isotope effect in the systems
with SSTs and quantitatively describing the experimental data.

This work is supported by the Russian Foundation for Basic
Research (projects 10-02-00140-a, 10-02-00598-a, 10-02-96011-a,
11-02-00708-a, 11-02-91335-NNIO-a, and 13-02-00374), by the Ural Branch of Russian Academy of Sciences through the young-scientist program, by the German projects DFG GR 1484/2-1, FOR 1346, by K\"oln University via German Excellence Initiative, and by the European network SOPRANO.


\begin{thebibliography} {99}
\bibitem{ISstate} As follows e.g. from the results of Ref.~\onlinecite{podlesnyak.prb11}, the magnetic states of Co ions generated by doping are most probably IS states. For the following treatment, however, it is not critical whether we deal with the IS or HS states.
\end{thebibliography}
\end{document}